\documentclass[aps,11pt,showpacs]{revtex4}
\usepackage{amssymb}
\usepackage{amsmath}
\usepackage{amsfonts}
\usepackage{natbib}
\begin{document}
\title{A closed-form necessary and sufficient condition for any two-qubit state to show hidden  nonlocality w.r.t the Bell-CHSH inequality}

\author{Rajarshi Pal }

\email{rajarshi@imsc.res.in}

\affiliation{Optics \& Quantum Information Group, The Institute of Mathematical
Sciences, C. I. T. Campus, Taramani, Chennai 600113, India}

\author{Sibasish Ghosh}

\email{sibasish@imsc.res.in}

\affiliation{Optics \& Quantum Information Group, The Institute of Mathematical
Sciences, C. I. T. Campus, Taramani, Chennai 600113, India }

\begin{abstract}
In this note, we discuss  a closed-form necessary and sufficient condition for any two-qubit state to show hidden nonlocality w.r.t the Bell-CHSH inequality. This is then used to numerically compute 
the relative  volume of states showing hidden Bell-CHSH non-locality , among all two-qubit states with one-sided reduction maximally mixed. 
\end{abstract}
\pacs{03.65.Bz,89.70.+c, 03.67.Mn, 42.50.Dv }
\maketitle

\section{Introduction:}
Nonlocality, other than being one of the most characteristic features of quantum mechanics  has also been established as a resource for quantum information processing(\cite{nonlocality-review}). Particularly, in recent years {\textit{device independent
quantum information processing}} has emerged where quantum nonlocality is considered to be  the  main resource as opposed to entanglement\cite{nonlocality-review}. The characterization and quantification of quantum non-locality is thus of prime importance  from an 
information theoretic point of view.

Nonlocality of certain quantum states can be {\textit{revealed}} by post-selection through local filters before performing a standard Bell-test. This phenomenon(called `hidden nonlocality') has received widespread attention(\cite{nonlocality-review},\cite{BHQ13}) in 
the study of quantum non-locality and its interrelation with entanglement  
ever since the first examples of it were produced in \cite{POP95}, \cite{Gis96}.However in spite of the progress made so far it is not known for any Bell inequality(in a closed-form), what are the necessary and sufficient conditions for a quantum state to show    
hidden non-locality. In this work we fill this gap by providing a closed-form necessary-sufficient condition for any two-qubit state to show hidden nonlocality with a single copy w.r.t the Bell-CHSH inequality.Or main result is given by Theorem 1 in the next section. 

\section{Hidden Bell-CHSH nonlocality}
{\textbf{Defn.}}

Consider  a local filtering transformation taking any two-qubit state $\rho$ to another two-qubit state
\begin{equation}
\rho'=\frac{(A \otimes B)\rho (A^{\dagger} \otimes B^{\dagger})}{Tr(A^{\dagger}A 
\otimes B^{\dagger}B\rho) } \label{local-filtering-eq}
\end{equation}
Then, $\rho$ is said to show hidden non-locality w.r.t the Bell-CHSH inequality  iff $\rho'$ violates the Bell-CHSH inequality \cite{CHSH} for at least one choice of $A$,$B$. 

Let  $R$ be the real $4 \times 4$ matrix  with $R_{ij}=Tr(\rho \sigma_i \otimes \sigma_j), i,j=0,1,2,3$ (where $\sigma_0=I_2 $). Further let $C_{\rho}=MRMR^T$ with  $M=\mbox{diag}(1,-1,-1,-1)$.

\vspace{5mm}
{\textbf{Theorem 1:}} Let $\lambda_i(C_{\rho}),(i=0,1,2,3)$ denote the eigenvalues of $C_{\rho}$ in descending order for an arbitrary two-qubit state $\rho$. Then, $\rho$ shows hidden 
 nonlocality w.r.t the Bell-CHSH inequality iff 
\begin{equation}
\lambda_1(C_{\rho}) + \lambda_2(C_{\rho}) > \lambda_0(C_{\rho}). 
\label{hid-nlc-cond}
\end{equation}
The maximum  Bell violation obtained from the   optimal filtered (or quasi-distilled) Bell-diagonal state being $2\sqrt{\frac{(\lambda_1(C_{\rho}) + \lambda_2(C_{\rho}))}{\lambda_0(C_{\rho})}}$.

\vspace{5mm}
{\textbf{Proof}}:

Under a local filtering transformation taking any two-qubit state $\rho$ to an unnormalized state \begin{equation}\rho'=(A \otimes B)\rho (A^{\dagger} \otimes B^{\dagger}), \label{local-filtering-eq0} \end{equation} 
the real $4 \times 4$ matrix $R$ transforms as \cite{Ver01}
\begin{equation}
\label{snlb-1}
R'\equiv Tr(\rho'\sigma_i \otimes \sigma_j)=L_A R L_B^T |det(A)||det(B)|
\end{equation}
with the Lorentz 
transformations $L_A$ and $L_B$ being given by,
\begin{eqnarray}
L_A= \frac{T(A \otimes A^*)T^{\dagger}}{|det(A)|}, \nonumber \\ 
L_B=  \frac{T(B \otimes B^*)T^{\dagger}}{|det(B)|},
\end{eqnarray}
and $ T= \frac{1}{\sqrt{2}} \begin{bmatrix} 1 & 0 & 0 & 1 \\ 0 & 1 & 1 & 0 \\ 0 & i & -i & 0 \\ 1 & 0 & 0 & -1  \end{bmatrix}$, with the  normalisation factor   $R'_{00}=Tr(\rho')= Tr(A^{\dagger}A 
\otimes B^{\dagger}B\rho)$.

{\textbf{Remark:}} Note that the filters $A$, $B$ must be of full rank for the Lorentz transformations to be finite.

It was  shown in \cite{Ver01} and \cite{Ver-lor} that by suitably choosing $A$ and $B$ and hence proper orthochronous Lorentz transformations $L_A$ , $L_B$
for {\textit{any}} $\rho$ we can have $R'$ to be either diagonal corresponding to a Bell-diagonal state $\rho'$ or of the  form,
\begin{equation}
 R'=R_{\rho'} = \begin{bmatrix} a & 0 & 0 & b \\ 0 & d & 0 & 0 \\ 0 & 0 & d & 0 \\ c & 0 & 0 & (b+c-a)   \end{bmatrix} 
\end{equation}
with   the corresponding $\rho'$(unnormalized) being 
\begin{equation}
\label{rho-after-local-filtering}
\rho'= \frac{1}{2 }\begin{bmatrix} b+c & 0 & 0 & 0 \\ 0 & a-b & d & 0 \\ 0 & d & (a-c) & 0 \\ 0 & 0 & 0 &  0   \end{bmatrix} .
\end{equation}
The  possible sets of real values of $b$, $c$ and $d$ are given by,
\begin{eqnarray}
&({\rm i})& b=c=\frac{a}{2}, \nonumber \\ 
&({\rm ii})& (d=0=c) \mbox{  and  }  (b=a),  \nonumber \\ 
&({\rm iii})& (d=0=b) \mbox{  and  } (c=a), \nonumber \\ 
&({\rm iv})& (d=0) \mbox{ and } (a=b=c) .  \label{value-cases}
\end{eqnarray}
Case (i) corrsponds to rank three or two states while the other cases corrrespond to either the product states $|00\rangle \langle 00|$ or the state $|0\rangle \langle 0| \otimes \frac{I}{2}$. 

\vspace{5mm}
From eqn. (\ref{snlb-1}) it follows that  the spectrum of $MR'MR'^T$ is given by
\begin{equation}
\label{c-matrix-transformation}
\lambda(MR'MR'^T) = |det(A)|^2  |det(B)|^2 \lambda(ML_ARL_B^TML_BR^TL_A^T) =|det(A)|^2  |det(B)|^2 \lambda(MRMR^T),   
\end{equation}
where we have used $L_A^TML_A=M=L_B^TML_B$. 
Now as the filters $A$ and $B$ are of full rank i.e, $det(A),det(B) \ne 0$ we have for each for each $i \in \{0,1,2,3\}$ 
\begin{equation}
\frac{\lambda_i({C_{\rho'}})}{\lambda_0(C_{\rho'})} =  \frac{\lambda_i({C_{\rho}})}{\lambda_0(C_{\rho})}. 
\label{eigfrac}
\end{equation}

Let us consider the following cases now.

\vspace{5mm}
(a)  $R'=\mbox{diag}(s_0,s_1,s_2,s_3)$.  $\rho'$  corresponds to a Bell-diagonal state which in turn violates the Bell-CHSH inequality (\cite{Hor95}) after normalization provided,
\begin{equation}
1< \frac{s_1^2}{s_0^2} + \frac{s_2^2}{s_0^2}= \frac{\lambda_1(C_{\rho'})
+ \lambda_2(C_{\rho'})}{\lambda_0(C_{\rho'})}=\frac{\lambda_1(C_{\rho})
+ \lambda_2(C_{\rho})}{\lambda_0(C_{\rho})}  \nonumber.
\end{equation}
(by eqn. (\ref{eigfrac})). This proves Thoerem 1 for this case.

\vspace{5mm}
(b) 
$\rho'$ is of the non Bell-diagonal form with $d \ne 0$ in eqn. (\ref{rho-after-local-filtering})  (case (i) of eqn. (\ref{value-cases})) .

\vspace{5mm}
It is easy to see by partial transposition that $\rho'$ must be entangled .

Further by using filters of the form of  
$A=\mbox{diag}(\sqrt{\frac{(a-c)}{(a-b)}}\frac{1}{n},1)$ and $B=\mbox{diag}(\frac{1}{n},1)$
we have, 
\begin{eqnarray}
\rho_1 &=& (A \otimes B) \rho (A^{\dagger} \otimes B^{\dagger}) \nonumber \\ 
&=& \frac{1}{2} \left(\frac{(b+c)(a-c)}{(a-b)n^4}| 00 \rangle \langle 00| + \frac{(a-c)}{n^2} (|01 \rangle \langle 01| + |10 \rangle \langle 10|) + 
\frac{d\sqrt{(a-c)}}{n^2\sqrt{(a-b)}} (|01 \rangle \langle 10| + |10 \rangle \langle 01|) \right) . 
\end{eqnarray}

By taking a very large positive no. $n$ , $\rho_2 = \frac{\rho_1}{Tr(\rho_1)}$ can be made to approach arbitrarily close to the Bell-diagonal state 
\begin{eqnarray}
\rho_3 &=& \frac{1}{2} ((|01 \rangle \langle 01| + |10 \rangle \langle 10|) + \frac{d}{\sqrt{(a-b)(a-c)}}(|01 \rangle \langle 10| + |10 \rangle \langle 01|)) \nonumber \\
&=& \frac{1}{4} ( I \otimes I + \frac{d}{\sqrt{(a-c)(a-b)}} \sigma_1 \otimes \sigma_1 + \frac{d}{\sqrt{(a-c)(a-b)}} \sigma_2 \otimes \sigma_2 - \sigma_3 \otimes \sigma_3 ).
\end{eqnarray}
Now,  from eqn. (\ref{rho-after-local-filtering})  we have $\lambda(C_{\rho'})= [(a-b)(a-c), (a-b)(a-c), d^2, d^2 ]$. 

From theorem 3 of ref. \cite{VW02} we also know that the optimal Bell-violation among the states connected to $\rho$ by local filtering transformations is obtained from the `quasi-distilled' state $\rho_3$. Hence by using eqn. (\ref{eigfrac}) 
we obtain an optimal Bell violation of amount

\begin{equation}
1 + \frac{d^2}{(a-b)(a-c)} =  \frac{\lambda_1(C_{\rho'})
+ \lambda_2(C_{\rho'})}{\lambda_0(C_{\rho'})} = \frac{\lambda_1(C_{\rho})
+ \lambda_2(C_{\rho})}{\lambda_0(C_{\rho})} > 1 
\end{equation}
(note that $(a-b)(a-c) \geq d^2$ by virtue of positivity of $\rho'$)

Thus states for which $\rho'$ is not Bell-diagonal ($d \ne 0$ case ) will {\textit{always}} violate the Bell-CHSH inequality after suitable local filtering transformation. 

\vspace{5mm}
(c)
$\rho'$ is of the non Bell-diagonal form with $d = 0$ in eqn. (\ref{rho-after-local-filtering}) (cases (ii), (iii) and (iv) of eqn. (\ref{value-cases})) . 
These states being of the product form must come from a separable $\rho$ (local filtering  with full rank filters being  invertible) and from eqns. (\ref{rho-after-local-filtering}) and (\ref{c-matrix-transformation}) we have $\lambda_i(C_{\rho})=\lambda_i(C_{\rho'})=0$ for all $i$. Thus Theorem 1 holds.  

\vspace{3mm}
Conversely, when eqn. (\ref{hid-nlc-cond}) is satisfied we can either filter or quasi-distill $\rho$ to a Bell-diagonal state with optimal Bell-violation $2\sqrt{\frac{(\lambda_1(C_{\rho}) + \lambda_2(C_{\rho}))}{\lambda_0(C_{\rho})}}$.

\hfill $\square$ 

\section{Applications}

Using theorem 1 we have numerically computed the relative  volume of states showing hidden Bell-CHSH non-locality , among all two-qubit states with one-sided reduction maximally mixed. The latter form a six parameter family isomorphic to the set of    
all qubit channels. The relative volumes of states which do {\textit{not}} show hidden Bell-CHSH non-locality and separable states turn out to be about $0.39$ and $0.24$ respectively, while that of states which satisfy the Bell-CHSH inequality {\textit{without}} post-selection
through local filters is about $0.81$ . Thus the post-selection restriction considerably reduces the difference between entangled and non-local states and it will be  interesting to see how far more it is reduced as one considers more inequalities like  
$I_{3322}$ \cite{Collins-Gisin-2003}.  

\section{Conclusion}
In this note  we have described a closed-form necessary and sufficient condition for any two-qubit state to show hidden nonlocality w.r.t the Bell-CHSH inequality.We believe this  is a useful step in the quantification of nonlocality and  will aid 
in further studies of quantum non-locality as a resource and in its comparison with entanglement. 
\bibliographystyle{plain}
\bibliography{qip} 

\begin{thebibliography}{10}

\bibitem{nonlocality-review}
Nicolas Brunner, Daniel Cavalcanti, Stefano Pironio, Valerio Scarani, and
  Stephanie Wehner.
\newblock Bell nonlocality.
\newblock {\em Rev. Mod. Phys.}, 86:419--478, Apr 2014.

\bibitem{CHSH}
John~F. Clauser, Michael~A. Horne, Abner Shimony, and Richard~A. Holt.
\newblock Proposed experiment to test local hidden-variable theories.
\newblock {\em Phys. Rev. Lett.}, 23:880--884, Oct 1969.

\bibitem{Collins-Gisin-2003}
Daniel Collins and Nicolas Gisin.
\newblock A relevant two qubit bell inequality inequivalent to the chsh
  inequality.
\newblock {\em Journal of Physics A: Mathematical and General}, 37(5):1775,
  2004.

\bibitem{Gis96}
N.~Gisin.
\newblock Hidden quantum nonlocality revealed by local filters.
\newblock {\em Physics Letters A}, 210(3):151 -- 156, 1996.

\bibitem{BHQ13}
Flavien Hirsch, Marco~T\'ulio Quintino, Joseph Bowles, and Nicolas Brunner.
\newblock Genuine hidden quantum nonlocality.
\newblock {\em Phys. Rev. Lett.}, 111:160402, Oct 2013.

\bibitem{Hor95}
R.~Horodecki, P.~Horodecki, and M.~Horodecki.
\newblock Violating bell inequality by mixed spin-12 states: necessary and
  sufficient condition.
\newblock {\em Phys. Lett. A}, 200(5):340--344, 1995.

\bibitem{POP95}
Sandu Popescu.
\newblock Bell's inequalities and density matrices: Revealing ``hidden''
  nonlocality.
\newblock {\em Phys. Rev. Lett.}, 74:2619--2622, Apr 1995.

\bibitem{Ver-lor}
Frank Verstraete, Jeroen Dehaene, and Bart De~Moor.
\newblock Lorentz singular-value decomposition and its applications to pure
  states of three qubits.
\newblock {\em Phys. Rev. A}, 65:032308, Feb 2002.

\bibitem{Ver01}
Frank Verstraete, Jeroen Dehaene, and Bart DeMoor.
\newblock Local filtering operations on two qubits.
\newblock {\em Phys. Rev. A}, 64:010101, Jun 2001.

\bibitem{VW02}
Frank Verstraete and Michael Wolf.
\newblock {Entanglement versus Bell Violations and Their Behavior under Local
  Filtering Operations}.
\newblock {\em Phys. Rev. Lett.}, 89(17):17--20, October 2002.

\end{thebibliography}

\end{document}